\documentclass[aps,prl,twocolumn,superscriptaddress,showpacs]{revtex4-1}

\usepackage{graphicx}
\usepackage{amsmath,amsfonts,amssymb}
\usepackage{bm}
\usepackage{color}

\newcommand{\didv}{{\textit{dI/dV}}}

\newcommand{\rsrt}{{$\sqrt{7}\times\sqrt{3}$}}

\begin{document}

\title{Imaging Josephson Vortices on the Surface Superconductor Si(111)-(\rsrt)-In using a Scanning Tunneling Microscope}

\author{Shunsuke Yoshizawa}
\affiliation{International Center for Materials Nanoarchitectonics, National Institute for Materials Science, 1-1 Namiki, Tsukuba 305-0044, Japan}

\author{Howon Kim}
\affiliation{The Institute for Solid State Physics, University of Tokyo, 5-1-5 Kashiwanoha, Kashiwa, Chiba Japan 277-8581, Japan}

\author{Takuto Kawakami}
\affiliation{International Center for Materials Nanoarchitectonics, National Institute for Materials Science, 1-1 Namiki, Tsukuba 305-0044, Japan}

\author{Yuki Nagai}
\affiliation{CCSE, Japan Atomic Energy Agency, Kashiwa, Chiba 277-8587, Japan}

\author{Tomonobu Nakayama}
\affiliation{International Center for Materials Nanoarchitectonics, National Institute for Materials Science, 1-1 Namiki, Tsukuba 305-0044, Japan}

\author{Xiao Hu}
\affiliation{International Center for Materials Nanoarchitectonics, National Institute for Materials Science, 1-1 Namiki, Tsukuba 305-0044, Japan}

\author{Yukio Hasegawa}
\affiliation{The Institute for Solid State Physics, University of Tokyo, 5-1-5 Kashiwanoha, Kashiwa, Chiba Japan 277-8581, Japan}

\author{Takashi Uchihashi}
\email{UCHIHASHI.Takashi@nims.go.jp}
\affiliation{International Center for Materials Nanoarchitectonics, National Institute for Materials Science, 1-1 Namiki, Tsukuba 305-0044, Japan}

\date{\today}

\begin{abstract}
We have studied the superconducting Si(111)-(\rsrt)-In surface using a $^3$He-based low-temperature scanning tunneling microscope (STM).
Zero-bias conductance (ZBC) images taken over a large surface area reveal that vortices are trapped at atomic steps after magnetic fields are applied.
The crossover behavior from Pearl to Josephson vortices is clearly identified from their elongated shapes along the steps and significant recovery of superconductivity within the cores.
Our numerical calculations combined with experiments clarify that these characteristic features are determined by the relative strength of the interterrace Josephson coupling at the atomic step.

\end{abstract}


\maketitle
The recent discovery of superconductivity in silicon surface reconstructions with metal adatoms was an unexpected surprise, because they are regarded as one of the thinnest two-dimensional (2D) materials ever possible \cite{Zhang_PbIn1ML,Uchihashi_InR7R3Super,Uchihashi_InR7R3Resistive,Yamada_InR7R3Magneto,Brun_PbSiDisorder}.
This class of surface 2D materials has now become relevant for extensive superconductor researches in progress \cite{Oezer_PbHardSuper,Qin_Pb2ML,Wang_SingleFeSe,Sekihara_1MLPb}.
Notably, these new studies have been advanced by surface analytical techniques such as scanning tunneling microscopy (STM) \cite{Zhang_PbIn1ML,Brun_PbSiDisorder,Qin_Pb2ML,Wang_SingleFeSe} and ultrahigh vacuum (UHV)-compatible transport measurement\cite{Uchihashi_InR7R3Super,Uchihashi_InR7R3Resistive,Yamada_InR7R3Magneto,Tegenkamp_1DPb,Yamazaki_R7R3}.

One ubiquitous feature of these surface systems is the presence of atomic steps.
Atomic steps are considered to strongly affect electron transport phenomena, because they potentially decouple neighboring surface terraces \cite{Crommie_StandingWave,Hasegawa_StandingWave,Uchihashi_1DIn,Matsuda_StepR}.
This could prevent superconducting currents from running over a long distance.
The presence of supercurrents through atomic steps has indeed been demonstrated by direct electron transport measurements\cite{Uchihashi_InR7R3Super,Uchihashi_InR7R3Resistive,Yamada_InR7R3Magneto}, and recent experiments indicated that atomic steps work as Josephson junctions \cite{Uchihashi_InR7R3Super,Brun_PbSiDisorder}.
Nevertheless, direct evidence of Josephson coupling has not been obtained yet, and possible local variation of its strength has remained an open issue.
This problem is also closely related to Josephson junctions formed at the grain boundaries in thin films of high-$T_c$ cuprates, which are of technological importance \cite{Kogan_JJThinFilm,Hilgenkamp_JJCuprate}.

In this Letter, we report on compelling evidence of the Josephson coupling at atomic steps on the surface superconductor Si(111)-(\rsrt)-In [referred to as (\rsrt)-In].
Zero-bias conductance (ZBC) images taken with a low-temperature (LT) STM reveal that vortices are present at atomic steps after magnetic fields are applied.
The crossover behavior from Pearl to Josephson vortices is evident from their characteristic elongated shapes and significant recovery of superconductivity within their cores.
This identification is strongly supported by our numerical calculations, which clarify their dependence on the interterrace Josephson coupling at the atomic step.

The experiment was performed using a UHV-LT-STM constructed at the Institute of Solid State Physics, University of Tokyo.
The STM head was accommodated within a $^3$He-based cryostat combined with a solenoid superconducting magnet, where magnetic field was applied in the normal direction to the sample surface \cite{Nishio_PbIsland}.
The temperature of the STM head $T_\mathrm{head}$ reaches below 0.5 K, which is sufficiently lower than the superconducting transition temperature $T_c\approx 3$ K of the (\rsrt)-In surface \cite{Zhang_PbIn1ML,Uchihashi_InR7R3Super,Uchihashi_InR7R3Resistive,Yamada_InR7R3Magneto}.
Samples were prepared by thermal evaporation of In onto a clean Si(111) substrate followed by annealing in UHV \cite{Zhang_PbIn1ML,Kraft_R7R3,Rotenberg_R7R3,Yamazaki_R7R3,Uchihashi_InR7R3Super,Uchihashi_InR7R3Resistive}.
Subsequently, the surface (\rsrt)-In structure was confirmed by reflection high energy electron diffraction (RHEED) and STM [for representative data, see Figs. 1(a)(b)].
The \didv spectra were recorded at a constant STM tip height in the ac lock-in detection mode by sweeping the sample bias voltage $V_\mathrm{s}$.
ZBC images were taken at $V_\mathrm{s}=0$ mV in the same mode after the feedback was stabilized at $V_\mathrm{s}=20$ mV at each pixel point.

\begin{figure}
\includegraphics[width=86mm]{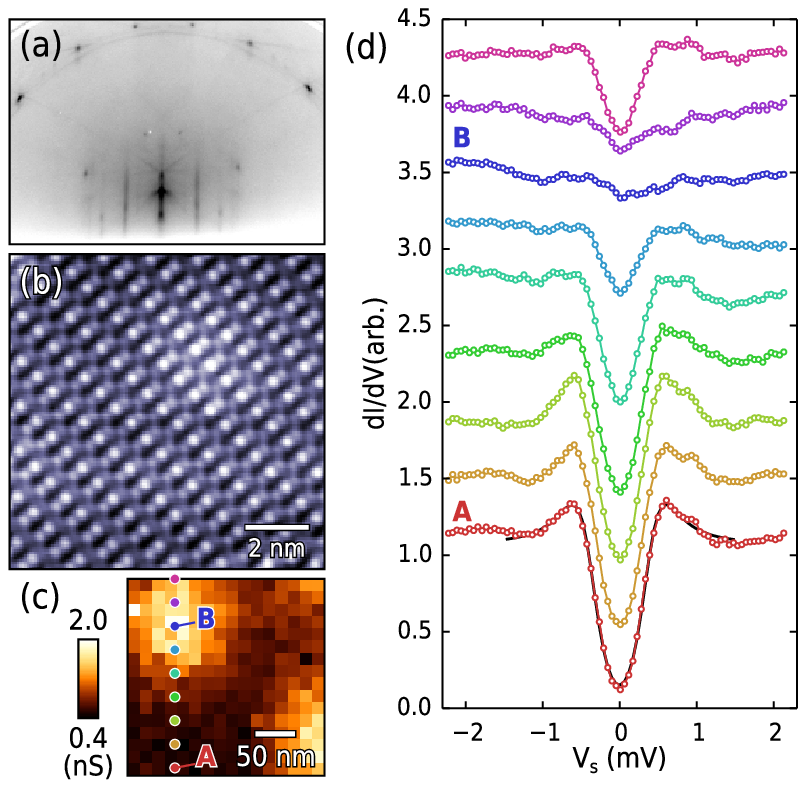}
\caption{(Color)
(a) Representative RHEED pattern of a (\rsrt)-In surface.  Electron beam energy: 2.5 keV.
(b) Representative STM image taken on a (\rsrt)-In surface. Set point: 500 mV, 50 pA.
(c) Zero-bias conductance (ZBC) image taken on a (\rsrt)-In surface at $T_\mathrm{head} < 0.5$ K and at $B_\mathrm{ext}=0.04$ T. Set point: 20 mV, 200 pA. Bias modulation: 610 Hz, $200 \mu \mathrm{V}$. The bright round features show Pearl vortex cores.
(d) Series of \didv spectra taken across the center of the left bright region. Set point: 20 mV, 600 pA. Bias modulation: 610 Hz, $50 \mu \mathrm{V}$. The curves are offset vertically for clarity. The locations for individual spectra are marked in the ZBC image in (c) in the same colors as used for spectral curves. The black curve is the result of fitting to the curve A using the Dynes formula.
}
\label{Fig1}
\end{figure}

First, we characterized our samples by measuring vortices on a flat area.
Figure 1(c) shows a ZBC image taken within a terrace of the (\rsrt)-In surface under a magnetic field of $B_\mathrm{ext}=0.04$ T.
The bright round regions (corresponding to high ZBC) indicate that vortices  were created due to the penetration of magnetic field \cite{Hess_vortex1,Tominaga_VortexSTM}.
Namely, while ZBC is low in the superconducting region due to the presence of the energy gap $\Delta$, it recovers towards the normal-state value as $\Delta$ is suppressed within the vortex core \cite{Tinkham_Textbook}.
To confirm this assignment, we obtained a series of site-dependent \didv spectra across the left bright feature [Fig. 1(d)].
At the location farthest from its center (marked as A), the \didv spectrum exhibited a characteristic superconducting energy gap structure with a dip around the zero bias and coherence peaks at $V_\mathrm{s} = \pm 0.60$ mV.
Our fitting analysis based on the Dynes formula with s-wave gap function \cite{Dynes_QPLifetime} gives an energy gap $\Delta = 0.39 \mathrm{meV}$, quasi-particle lifetime broadening $\Gamma = 0.00 \mathrm{meV}$, and the sample temperature $T_\mathrm{sample} = 1.3$ K.
\footnote{
Energy gap $\Delta$ obtained here is smaller than $\Delta=0.57$ meV reported previously for this surface \cite{Zhang_PbIn1ML}.
This may be due to the residual disorder found in the present sample.
}
(see the black line overlapped on Curve A).
As the spectral site approached the center (marked as B), the zero-bias dip and the coherence peaks were both strongly suppressed, indicating breaking of superconductivity.
We note that the vortices found here should be called Pearl vortices (PVs) because the present system consists of an atomically thin 2D superconductor
\cite{PearlVortex, Tafuri_PearlVortex}.
\footnote
{For a 2D superconductor with a thickness $d$, the characteristic length governing the magnetic field distribution is given by Pearl length $\Lambda=2\lambda^2/d$, where $\lambda$ is London penetration depth. The vortices interact with each other like $E_\mathrm{int} \propto \log r$  as long as $r<\Lambda$. The vortex is then called the Pearl vortex instead of the Abrikosov vortex in a three-dimensional (3D) superconductor, but their core structures are essentially the same. The magnetic flux size of a Pearl vortex is given by $\Lambda$, which is estimated to be as large as 4.4 mm here. Since the magnetic field distribution is considered to be uniform, it does not affect the structure of a vortex core (See Sec. 1 of Supplemental Material [url], which includes Ref.~\cite{Raychaudhuri_InSuper}).
}
For the following images, ZBC is normalized by the \didv value at a coherence peak at each pixel point to enhance the signal-to-noise ratio.

\begin{figure*}
\includegraphics[width=160mm]{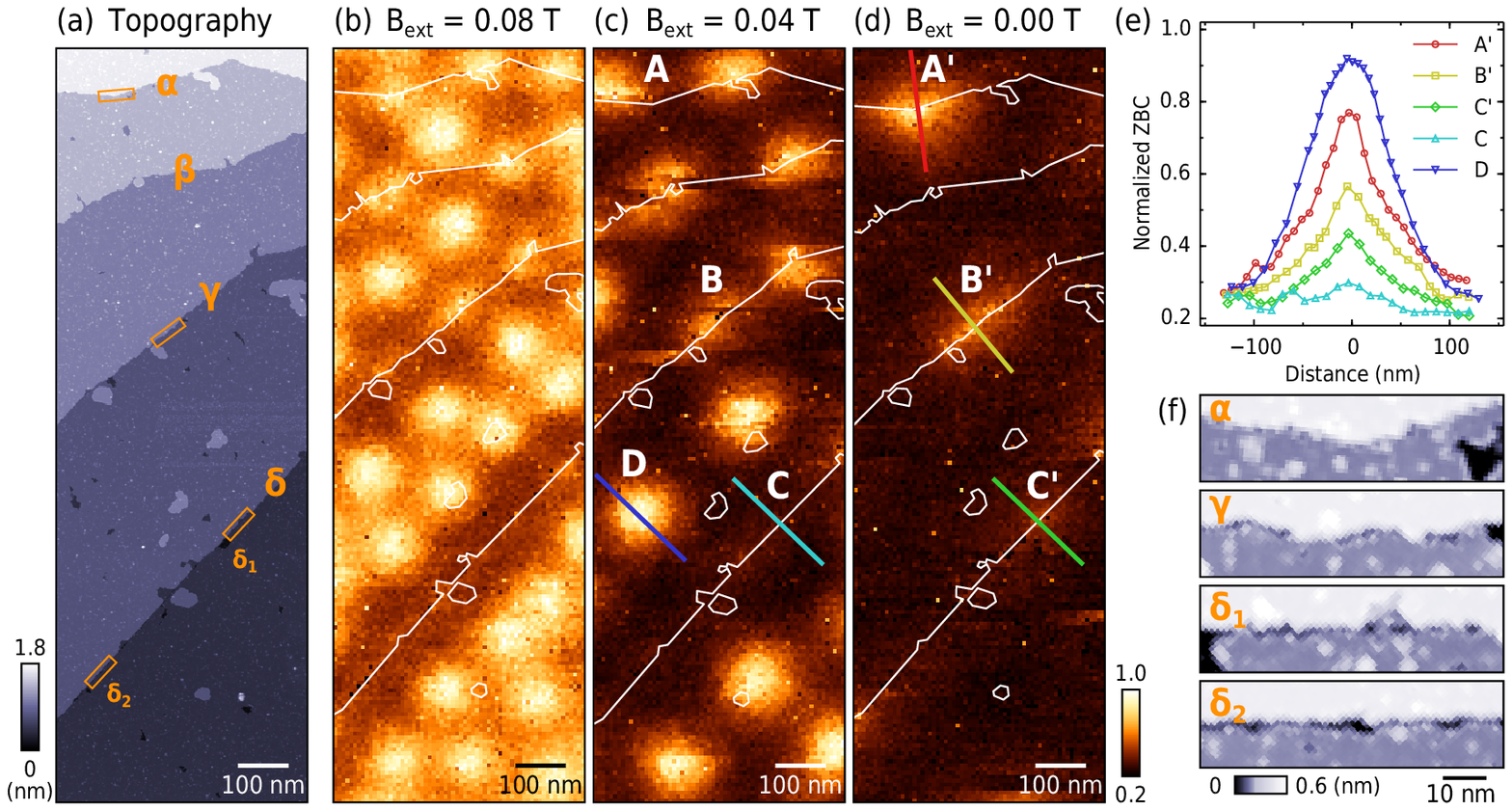}
\caption{(Color)
(a) Large-scale STM image of a (\rsrt)-In surface, where the terraces are separated by atomic steps marked as $\alpha$, $\beta$, $\gamma$, and $\delta$ from top to bottom. Set points: 90 mV, 10 pA.
(b)-(d) ZBC images of the same area as shown in (a) taken under different magnetic fields $B_\mathrm{ext}$: (b) $B_\mathrm{ext}=0.08$ T, (c) $B_\mathrm{ext}=0.04$ T, (d) $B_\mathrm{ext}=0$ T. Set point: 20 mV, 200 pA. Bias modulation: 610 Hz, $200 \mu \mathrm{V}$. The positions of the atomic steps are depicted by thin solid lines.
(e) Spatial profiles of ZBC plotted along the thick solid lines shown in (c)(d), which are indicated by the nearby markers A', B', C', C, and D.
(f) Magnified topographic images at steps $\alpha$, $\gamma$, and $\delta$ cut from the regions marked by the rectangles in (a).
}
\label{Fig2}
\end{figure*}

Further experiments on wider surface regions allowed us to access more details of vortices in the present system.
Figure 2(a) shows an STM topography image with an area of 500 nm $\times$ 1500 nm.
The surface consists of flat terraces separated by steps with the single atomic height of 0.31 nm, which are indicated as $\alpha$, $\beta$, $\gamma$, and $\delta$ from top to bottom.
ZBC images were taken on the same area under different magnetic fields of $B_\mathrm{ext}=0.08, 0.04, 0$ T in this order, as displayed in Figs. 2(b)-(d).
The locations of the atomic steps are designated by thin solid lines.
At $B_\mathrm{ext}=0.08$ T, PVs with bright round features formed a closely packed triangular lattice within each terrace.
Reduction of magnetic field to $B_\mathrm{ext}=0.04$ T decreased the number of vortices on terraces as expected.

When the magnetic field was set to zero, vortices disappeared from the terraces, but slightly bright regions remained at some points along the steps [Fig. 2(d)].
Note that similar features were also present along the steps at finite fields [Figs. 2(b)(c)].
They are not simply regions where superconductivity is suppressed due to the presence of steps or disorder nearby.
This is evident from the fact that the features change their positions under different magnetic fields, as seen from comparison of features A and A'.
Similarly, comparison of regions C and C' shows that ZBC increased at this location [see Fig. 2(e) for the ZBC profiles].
Furthermore, a sudden change in contrast is visible near feature B, indicating that it is mobile even under a constant field.
The above observations clearly show that these bright features are vortices trapped at the atomic steps.

The vortices at steps are anomalous when compared to the PVs on terraces.
Here we focus on vortices A', B', and C' in Fig. 2(d).
First, their shapes are elongated along the steps as seen from vortices A' and B';
the full width at half maxima (FWHM) along and across the step are 162 and 80 nm for vortex A', and 213 an 103 nm for vortex B'.
\footnote{
See Sec. 2 of Supplemental Material [url].
}
Vortex C' is largely spread along the step and appears to be disturbed by defects and/or temporal fluctuations.
In contrast, PVs are isotropically round as seen from vortex D in Fig. 2(c), with a FWHW of $94\pm 5$ nm.
Second, ZBC values measured at the centers are lower than those for PVs.
This is quantitatively depicted in Fig. 2(e) as the ZBC profiles taken along the thick lines across vortices A', B', C', and D.
It means that the superconducting energy gap at the core recovers towards the zero-field value, while there is essentially no energy gap for a PV \cite{Hess_vortex1,Tinkham_Textbook}.
As explained below, these anomalies are the direct consequences of crossover to Josephson vortex (JV) and show that the atomic steps work as Josephson junctions.
\footnote{
We stress that these observations were made possible through the STM measurement.
	Previous studies on JVs using scanning superconducting quantum interference devices on cuprates detected magnetic field distribution but did not access information on the vortex cores \cite{Hilgenkamp_JJCuprate}.
}
\begin{figure}
\includegraphics[width=80mm]{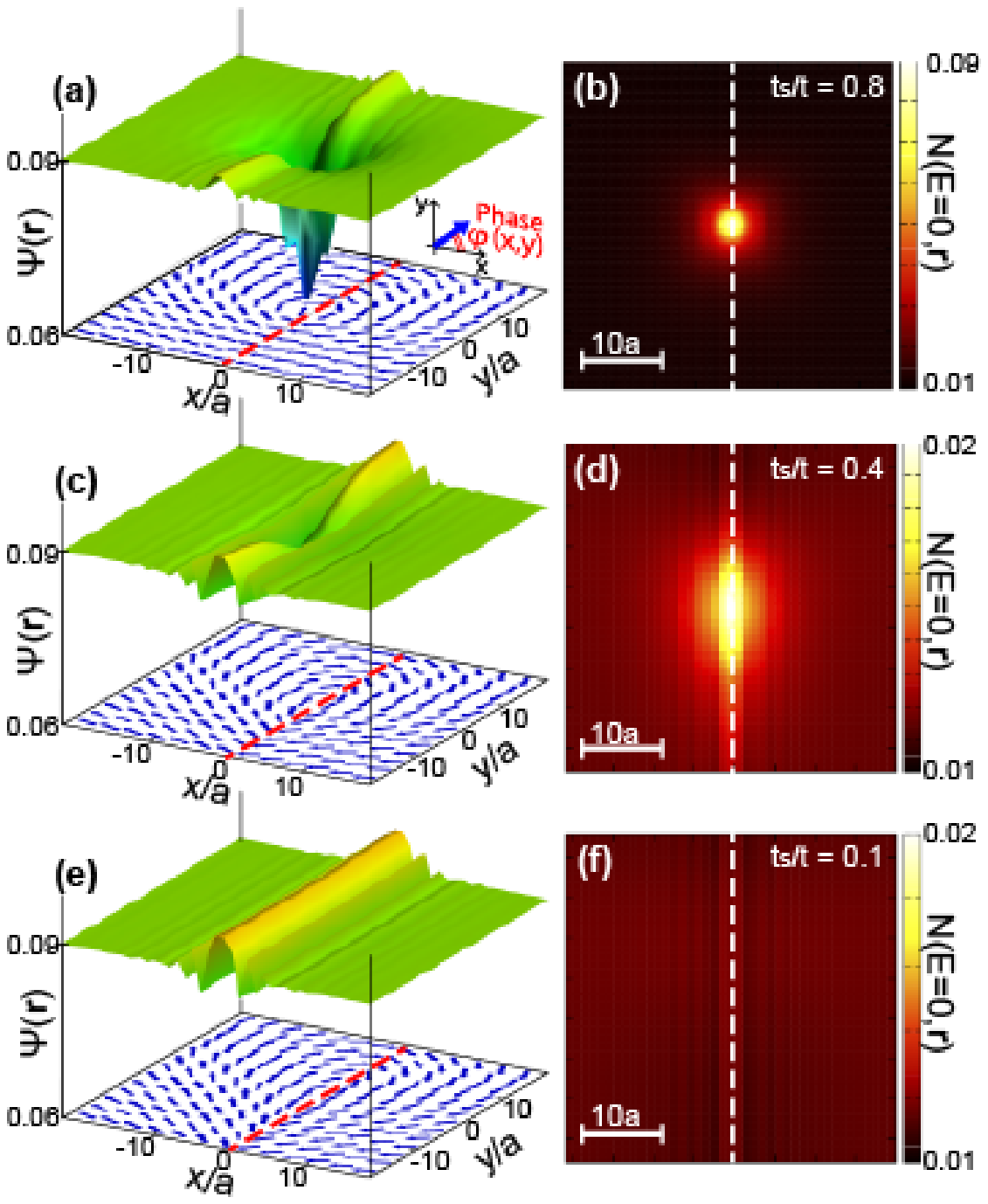}
\caption{(Color)
Numerically obtained spatial profile of the order parameter $\Psi(\bm{r})$ [(a)(c)(e)] and the zero energy density of state ${N}(E\!=\!0,\bm{r})$ [(b)(d)(f)]. The direction of an arrow in (a)(c)(e) denotes the phase $\phi(x,y)$ of the order parameter. The dashed lines indicate the place where the Josephson coupling was modeled as a reduced hopping strength $t_\mathrm{s}$. The length scale for $x$ and $y$ is the lattice constant $a$. Results in (a)(b), (c)(d), and (e)(f) are for hopping strength $t_\mathrm{s}/t=0.8,\ 0.4, \ 0.1$. We set the other parameters $\mu\!=\!-2.5t$ and $V\!=\!-3.0t$.
}
\label{Fig3}
\end{figure}

Suppose that a vortex is created by penetration of magnetic field through a Josephson junction line and its surrounding region.
Here the phase evolution due to supercurrent circulation around the core includes phase shifts $\Delta\phi$ at Josephson junctions.
In the simplest case, $\Delta\phi$ is related to the supercurrent density $J_s$ through the following relation \cite{Tinkham_Textbook}:
\begin{equation}
J_s = J_c \sin \Big[\Delta\phi-(2\pi/\Phi_0)\int \mathbf{A}(\bm{s}) \cdot d\mathbf{s} \Big],
\label{eq:Josephson_relation}
\end{equation}
where $J_c$, $\Phi_0$, and $\int \mathbf{A}(\bm{s}) \cdot d\mathbf{s}$ denote the critical current density of the Josephson junction, magnetic flux quantum ($=h/2e$), and path integral of vector potential at the junction, respectively.
This leads to two important properties regarding the vortex \cite{Blatter_VortexReview}.
First, the circulation of supercurrent near the center is strongly deformed and the vortex core is elongated along the junction line by a factor of $(J_c/J_0)^{-1}$, where $J_0 (> J_c)$ is the critical current density in the superconducting regions.
Second, the breaking of superconductivity around the core is weakened as $J_c/J_0$ decreases.
The amplitude of the superconducting order parameter at the center $|\Psi_\mathrm{center}|$ is given by
\begin{equation}
|\Psi_\mathrm{center}| \approx \left[1-(J_c/J_0)^2 \right] |\Psi_0|,
\label{eq:order_parameter_JV}
\end{equation}
where $|\Psi_0|$ is the order parameter in the absence of magnetic field and supercurrent.
The vortex should be called a JV when the supercurrent distribution near the junction line is nearly parallel and the suppression $\Delta |\Psi_\mathrm{center}|\equiv |\Psi_0|-|\Psi_\mathrm{center}| \approx (J_c/J_0)^2 |\Psi_0|$ is sufficiently smaller than $|\Psi_0|$.
This terminology is consistent with the common usage of JVs in layered superconductors, which are created by magnetic field parallel to the layers \cite{Koshelev_JVReview,Moll_TranstionAVtoJV}.
\footnote
{In a 3D system, the transition from Abrikosov to Josephson vortex occurs when the elongated core size exceeds the London penetration depth $\lambda$ [A. Gurevich, Phys. Rev. B \textbf{46}, 3187 (1992)]. In the present 2D system, vortices at junctions are quite different since the problem involves a nonlocal equation as opposed to the local sine-Gordon equation in the 3D case \cite{Kogan_JJThinFilm}. Hence this definition is not applicable here.
}

To compare the theoretical prediction with our experiment more directly, we numerically calculated the order parameter and the density of states (DOS) using the Bogoliubov-de Gennes (BdG) equation for a 2D tight-binding model:
\begin{eqnarray}~\label{eq:tight-bdg}
	\sum_{j}\left(\begin{array}{cc}
	\hat{K}_{i,j} &\hat{\Delta}_{i,j}\\
	\hat{\Delta}^\ast_{j,i} & -\hat{K}^\ast_{i,j}
	\end{array}\right)
	\left(\begin{array}{c}
	u_\gamma(\bm{r}_j)\\ v_\gamma(\bm{r}_j)
	\end{array}\right)
	=E_{\gamma}
	\left(\begin{array}{c}
	u_\gamma(\bm{r}_i)\\ v_\gamma(\bm{r}_i)
	\end{array}\right).
\end{eqnarray}
The single particle part is given by $\hat{K}_{i,j}=-t_{ij}\exp\left[i(\pi/\Phi_0)\int_{\bm{r}_i}^{\bm{r}_j}\bm{A}(\bm{s})\cdot d\bm{s}\right]-\mu\delta_{ij}$ with $t_{ij}$ the hopping strength.
The Josephson junction was modeled as a straight line with one atomic spacing where the hopping strength $t_{\mathrm{s}}$ is reduced from a constant hopping strength $t$ elsewhere.
Then the Josephson parameter $J_\mathrm{c}/J_0$ is represented by the ratio $t_\mathrm{s}/t$ according to Ambegaokar-Baratoff's equation \cite{Ambegaokar-Baratoff}.
Equation~(\ref{eq:tight-bdg}) was solved self-consistently ~\cite{Covaci2010,Nagai2012,NagaiPRB} to obtain the pair potential $\Delta(\bm r_{i})=\hat{\Delta}_{i,j}=\delta_{ij}V\sum_{\gamma}u_{\gamma}(\bm{r}_i)v_{\gamma}(\bm{r}_j)f(E_{\gamma})$ and DOS ${N}(E,\bm{r}_i)=\sum_{\gamma}|u_{\gamma}(\bm{r}_i)|^2\delta(E-E_{\gamma})$.
\footnote{
See Sec. 3 of Supplemental Material [url], which includes Ref.~\cite{Takigawa_NMRVortex}.
}

Figures 3(a)-(f) display the order parameter $\Psi(\bm r)=\Delta(\bm r)/V$ [(a)(c)(e)] and zero-energy DOS ${N}(E=0,\bm{r})$ [(b)(d)(f)] calculated for $t_\mathrm{s}/t=0.8, \ 0.4, \ 0.1$.
For $\Psi(\bm r)$, its amplitude $|\Psi(\bm r)|$ and phase $\phi(\bm r)$  are shown in the upper and lower panels within each figure, respectively.
The location of the Josephson coupling line (where $t_{ij}=t_\mathrm{s}$) is indicated by the dashed lines.
While the suppression of $|\Psi(\bm r)|$ is strong and the spatial distribution of $\phi(\bm r)$ is almost cylindrically symmetric for $t_\mathrm{s}/t=0.8$, the former becomes weaker and the latter is elongated along the junction line as $t_\mathrm{s}/t$ is reduced to 0.4 and 0.1.
Accordingly, the characteristics of ${N}(E=0,\bm{r})$ are changed; its magnitude around the center is decreased as $t_\mathrm{s}/t$ is reduced, while the spatial distribution becomes strongly elliptic.
Considering that ZBC is proportional to DOS, this evolution directly corresponds to the observed changes for vortices A', B', and C' in Fig. 2(d).
Thus the coupling strength $J_c$ at steps $\alpha,\gamma,\delta$ decreases in this order.
From the comparison of the experiment and the theory, $J_c/J_0$ is estimated to be $\sim 0.4$ for step $\gamma$ where vortex B' is located.
Step $\delta$ has a weak coupling $J_c/J_0 \ll 0.4$ and, according to the above definition, vortex C' can be safely called a JV.
We estimate $J_c=1.8 \ \mathrm{A/m}$ from the previous macroscopic transport measurement \cite{Uchihashi_InR7R3Super} and $J_0=19-62 \ \mathrm{A/m}$ from the present study, leading to $J_c/J_0=0.029-0.095$.
\footnote{
See Sec. 4 of Supplemental Material [url], which includes Ref.~\cite{Raychaudhuri_InSuper}.
}
This justifies our theoretical analysis because $J_c$ determined above should reflect the weakest interterrace coupling, being consistent with $J_c/J_0 \ll 0.4$ at step $\delta$.

The differences in $J_c/J_0$ clarified above may be attributed to the local atomic-scale structures along the steps.
Figure 2(f) shows topographic images near steps $\alpha$, $\gamma$, $\delta$ where vortices A', B', C' are located [marked by the rectangles in Fig. 2(a)].
Grooves are visible along step $\delta$, indicating that the superconducting indium layers did not grow up to the step edge.
This should result in a weak electronic coupling between the upper and lower terraces \cite{Uchihashi_1DIn} and hence in a low $J_c/J_0$.
In contrast, such a structure is nearly absent for step $\alpha$, which helps to establish a stronger interterrace coupling.

Finally, we remark on possible JVs in Fig. 2(b) under a high magnetic field. All visible bright features in the image counts for number of vortices $N_\mathrm{vis}=26$, which is different from $N_\mathrm{theory}= B_\mathrm{ext}S/\Phi_0=29$ (imaging area $S=500\mathrm{nm} \times 1500\mathrm{nm}$, $B_\mathrm{ext}=0.08$ T, $\Phi_0= 2.07\times 10^{-15} \mathrm{Tm^2}$). The missing flux quanta are $N_\mathrm{theory}-N_\mathrm{vis} = 3$ and they should exist as JVs along step $\delta$.

In conclusion, we have observed the crossover from PV to JV at atomic steps on the (\rsrt)-In surface by taking ZBC images using a LT-STM.
The present work provides compelling evidence and local information for Josephson coupling at atomic steps.

This work was financially supported by JSPS under KAKENHI Grants No. 25247053, No. 25286055, No. 25400385, No. 24340079 and by World Premier International Research Center (WPI) Initiative on Materials Nanoarchitectonics, MEXT, Japan.
The calculations was performed using the supercomputing system PRIMERGY BX900 at the Japan Atomic Energy Agency.


\bibliographystyle{apsrev4-1} 
\bibliography{MyEndNoteLibrary}

\end{document}